# Optimization of a multi-TW few-cycle 1.7-µm source based on Type-I BBO dual-chirped optical parametric amplification

LU XU,[1,4] KOTARO NISHIMURA,[1,2] YUXI FU,[1,3] AKIRA SUDA,[2] KATSUMI MIDORIKAWA,[1] AND EIJI J. TAKAHASHI[1,*]

[1]*Extreme Photonics Research Group, RIKEN Center for Advanced Photonics, RIKEN, 2-1 Hirosawa, Wako, Saitama 351-0198, Japan*
[2]*Department of Physics, Tokyo University of Science, 2641 Yamazaki, Noda, Chiba 278-8510, Japan*
[3]*Current address: Xi'an Institute of Optics and Precision Mechanics, Chinese Academy of Sciences, Xi'an, Shangxi 710119, China*
[4]*lu.xu@riken.jp*
*\*ejtak@riken.jp*

**Abstract:** This paper presents the optimization of a dual-chirped optical parametric amplification (DC-OPA) scheme for producing an ultrafast intense infrared (IR) pulse. By employing a total energy of 0.77 J Ti:sapphire pump laser and type-I BBO crystals, an IR pulse energy at the center wavelength of 1.7 µm exceeded 0.1 J using the optimized DC-OPA. By adjusting the injected seed spectrum and prism pair compressor with a gross throughput of over 70%, the 1.7-µm pulse was compressed to 31 fs, which resulted in a peak power of up to 2.3 TW. Based on the demonstration of the BBO type-I DC-OPA, we propose a novel OPA scheme called the "dual pump DC-OPA" for producing a high-energy IR pulse with a two-cycle duration.

## 1. Introduction

A variety of applications are increasingly supported by ultrashort and high-power coherent extreme ultraviolet (XUV)/soft-X-ray sources, such as biological imaging by "water window" (280–530 eV) soft X-rays [1], high spatial resolution microscopy [2], the study of electron dynamics by attosecond XUV/soft-X-ray pulses [3], and others [4, 5]. Currently, high order harmonic generation (HHG) driven by femtosecond lasers is an efficient way of generating ultrashort coherent XUV/soft-X-ray pulses. To scale up the output photon flux of HHG for practical applications, a method that uses loosely focused geometry [6, 7] with a high-energy laser pulse has been employed [8–10]. With this loosely focused geometry, the HHG flux can achieve a micro-joule class in the XUV region [11]. The maximum photon energy of HHG is also an essential feature in terms of practical applications. According to the "cut-off law" of HHG, which can be explained in terms of the three-step model [12], in the interaction between an intense laser field and atoms or molecules using a semiclassical model, the maximum photon energy of HHG is given by $I_p+3.17U_p$, where $I_p$ is the ionization potential and $U_p = 9.38 \times 10^{-14} I[W/cm^2](\lambda[\mu m])^2$ is the ponderomotive energy. This means that the ponderomotive energy is proportional not only to the driving laser intensity ($I$) but also to the square of the wavelength of the laser ($\lambda^2$). Thus, combining a high-energy infrared (IR) femtosecond laser source and loosely focused geometry of HHG achieves both photon flux scaling [6] and photon energy scaling [13] simultaneously.

Currently, optical parametric chirped-pulse amplification (OPCPA) is commonly employed to produce ultrafast high-energy IR pulses [14]. A pump laser with pulse duration of picosecond level is required to maintain efficient broadband amplification in the IR OPCPA process, as the picosecond level pump pulses with a larger bandwidth than standard nanosecond pump pulses and can match a greater broadband spectral range of signal pulses [15]. Therefore, the development of IR OPCPA strongly depends on the progress of the picosecond pump laser.

A dual-chirped optical parametric amplification (DC-OPA) method, which is a variant of the OPCPA scheme, was theoretically proposed in 2011 [16] for the production of high-energy IR pulses. DC-OPA has certain unique characteristics [17] that OPCPA does not have. For example, in DC-OPA, the pump and seed are produced from a single laser system and are stretched to eliminate damage to nonlinear crystals, and a good phase matching (PM) between the pump and seed pulses is obtained by manipulating the chirp combination of pump and seed pulses. DC-OPA produces multi-TW IR pulses in the wavelength region from 1 to 3 µm [18–20]. However, its pulse duration is limited to the multi-cycle regime due to the narrowband PM of DC-OPA.

In this paper, the generation of broadband IR laser pulses of over 0.1 J at a wavelength of 1.7 µm with a repetition rate of 10 Hz based on the DC-OPA scheme pumped by a joule-class Ti:sapphire laser is demonstrated. A pulse compression to 31 fs duration (5.5 cycles) was achieved after the DC-OPA yielding a peak power of 2.3 TW. In addition, a numerical model formulated in the frequency domain, in which the effects of non-collinear configuration, temporal and spatial walk-off, dispersion, and diffraction are all considered [20], was extended for analysis and optimization of the IR DC-OPA laser system. Here, the injected parameters in the simulation were consistent with the corresponding experiment. The optimal chirp combination between the pump and seed pulses, the influence of third-order dispersion (TOD) on amplification, and the spectral phase distortion caused by the saturation effect in the OPA process were numerically simulated and experimentally demonstrated. Moreover, a novel dual pump DC-OPA scheme, which overcomes the narrow bandwidth of the Type-I BBO DC-OPA at 1.7 µm, was demonstrated in our simulations. The dual pump DC-OPA solved the bottleneck in the bandwidth of the DC-OPA and was capable of producing a two-cycle pulse duration at a center wavelength of 1.7 µm.

## 2. Experimental setup

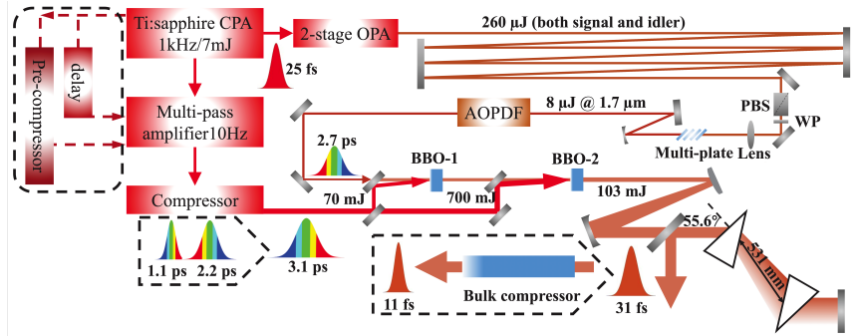

Fig. 1. Multi-TW few-cycle IR source based on DC-OPA; AOPDF, acousto-optic programmable dispersive filter, PBS, polarizing beam-splitter, WP, wave plate.

The detailed experimental setup is illustrated in Fig. 1 and consisted of a 1 kHz front-end Ti:sapphire preamplifier and a 10 Hz back-end power amplifier. The output pulse at the front-end preamplifier was partially split into two beams. The uncompressed pulses with an energy of 1.4 mJ and pulse duration of ~150 ps were further amplified to 1.1 J by a 10 Hz back-end multi-pass power amplifier. Subsequently, a grating compressor with a throughout efficiency of 70% was employed to manipulate both the chirp sign and value of the compressed pulses, which produced an optimized pump pulse for the DC-OPA. Finally, pulses with an energy of 0.77 J, beam diameter of 20 mm, and pulse duration of ~3.1 ps (with up-chirp) were obtained as pump pulses for a two-stage IR DC-OPA. At the same time, the compressed laser pulses (4 mJ, 25 fs) from a 1 kHz front-end preamplifier were used to pump a collinear two-stage OPA, which generated IR pulses with wavelength tunability in the range of 1.2–1.6 µm for the signal pulses and 1.6–2.6 µm for the idler pulses. In this experiment, the central wavelength of the 1.7-µm idler pulses was used as seed pulse for the DC-OPA. Thanks to the difference-frequency generation scheme, the carrier-envelope phase of the seed pulse was passively stabilized. The seed pulses propagated paths of several meters to compensate for the time delay to the pump laser pulses (0.8 µm). As the bandwidth of the seed pulses from the OPA was not sufficient to produce a few-cycle duration, the bandwidth was broadened by focusing (where the focus length of the lens was 100 mm) on a multi-plate setup [21] with an intensity of ~$1 \times 10^{13}$ W/cm$^2$. The multi-plate setup consisted of four fused silica plates oriented under Brewster's angle, with thicknesses of 0.1, 0.2, 0.3, and 0.5 mm in sequence. By optimizing the separations between the plates [22], the injected seed pulses were spectrally broadened (corresponding to a transform-limited pulse duration of 15 fs) with a pulse energy of 8 µJ, as shown in Fig. 2(a). Subsequently, an acousto-optic programmable dispersive filter (AOPDF; FASTLITE, Inc) was utilized to control the pulse duration of the broadened laser pulses at a wavelength of 1.7 µm with a precisely manipulated dispersion (down-chirp) before seeding the first stage of the DC-OPA. The DC-OPA element was constructed in a two-stage configuration, where type-I BBO crystals with a cutting angle (θ) of 20° were employed in both DC-OPA stages, and the thicknesses of the crystals in the first and second stages were 4 and 3 mm, respectively. A broadband dielectric-coated beam splitter was used to reflect 70 mJ pulse energy to pump the first stage of the DC-OPA and transmit 700 mJ pulse energy to pump the second stage of the DC-OPA. To maintain nearly the same pump intensity for both stages, the pump beam size for the first stage was down-collimated by a factor of ~3.2. After the first stage, the signal pulses with a beam diameter of 5 mm were extracted and expanded to a dimeter of ~ϕ 20 mm to match the beam size of the pump laser pulses in the second stage. The amplified signal pulses from the second stage of the DC-OPA were then temporally compressed by a water-free fused-silica (OHARA, SK-1310) prism pair compressor.

### 3. Experimental results

*3.1 PM tailoring of Type-I BBO DC-OPA*

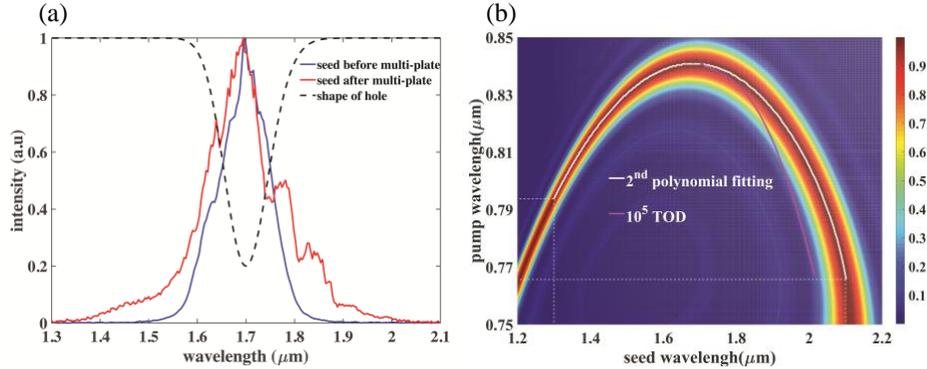

Fig. 2. (a) Seed spectrum evolution. Blue solid line, the seed spectrum before multi-plate setup; red solid line, the seed spectrum after multi-plate setup; black dashed line, the shape of the hole used in the injected seed for DC-OPA. (b) The calculated PM efficiency as a function of pump and signal wavelengths. White solid line, the 2$_{nd}$ polynomial fitting of ideal chirp matching between pump and signal pulses; pink solid line, the chirp matching with signal pulses with an added 10$_5$ TOD value.

The phase matching (PM) efficiency [23], sinc$_2$($\Delta kL/2$), where $\Delta k$ represents the phase mismatch and $L$ is the BBO crystal length, was plotted as a function of the pump and signal wavelengths in Fig. 2 (b). It was predicted that the opposite chirp combination between the pump and signal would provide larger mismatch tolerance around the signal center wavelength

(1.7 µm). Under the above circumstances, the opposite chirp matching was used in the experiment. The PM character on Type-I BBO DC-OPA showed that the broadband 1.7 µm DC-OPA for supporting a few-cycle duration would not work. Later, we discuss the dual pump DC-OPA system as a means of overcoming this technical obstacle.

*3.2 Optimization and pulse compression of Type-I BBO DC-OPA*

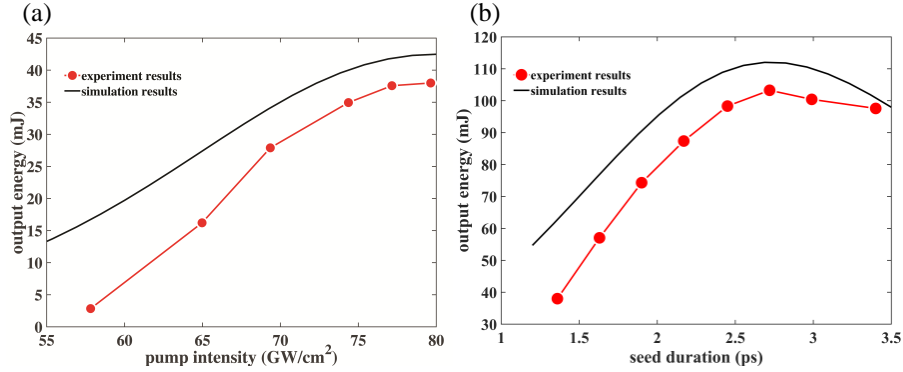

Fig. 3. The output energy after two stages of DC-OPA. (a) Energy scaling as a function of pump intensity. (b) Energy evolution as a function of seed pulse duration. Both in (a) and (b), the black solid line and red solid circles are simulation results and experiment results, respectively.

First, we performed an experiment to study the evolution of the output energy of signal pulses in order to achieve the optimal chirp combination (or pulse duration matching) between the pump and seed pulses. Here, the seed pulse duration was optimized up to group delay dispersion (GDD) without TOD in each case. Fig. 3(a) shows the output pulse energy after two stages of the DC-OPA as a function of the pump intensity when the seed and pump pulse durations were fixed to 1.36 ps (GDD, +9000 fs$_2$) and 3.1 ps (GDD, −30000 fs$_2$), respectively. As the pump intensity increased from 58 to 79 GW/cm$_2$ (total pump energy increased from 0.57 to 0.77 J), the output energy grew to 38 mJ, where it reached the maximum value from the trend predicted by both the experiment and simulation results. To optimize the timing overlap between the pump and signal pulses, we increased the signal pulse duration from 1.36 to 3.4 ps, and the corresponding output energy evolution is shown in Fig. 3(b). An output energy of 103 mJ was obtained with a signal pulse duration of 2.7 ps and decreased with a signal pulse duration of over 2.7 ps. As shown in Fig. 3(b), the trend demonstrated in the experiment was in good agreement with the simulation prediction.

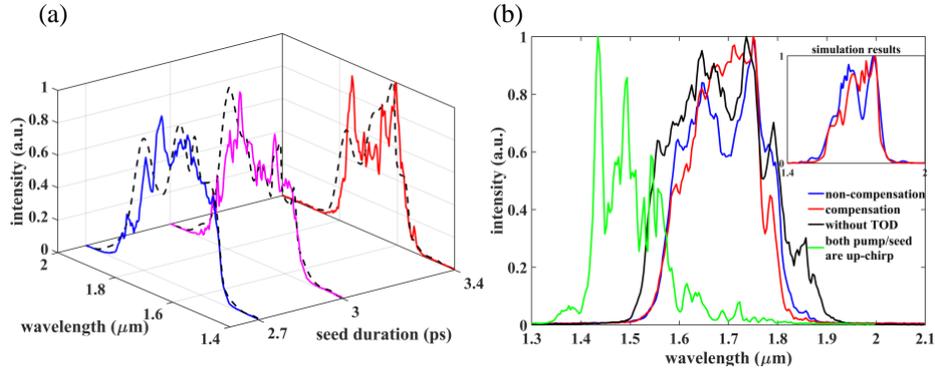

Fig. 4. (a) The output spectrum evolution with a seed pulse duration of 2.7 ps (blue solid line), 3.0 ps (pink solid line), and 3.4 ps (red solid line). Black dashed lines, simulation results. (b) The output spectrum with the influence of TOD and saturation effect in the OPA process. Black solid line, without TOD influence; blue solid line, with the TOD caused by the prism pair

compressor and saturation effect in the OPA process; red solid line, with the TOD caused by the prism pair compressor and no saturation effect in the OPA process; green solid line, the output spectrum with the same chirp sign in pump and seed; the corresponding simulation results are also shown in the inset of (b).

To explain the feature of energy evolution as a function of seed pulse duration, the output spectrum evolution both in the experiment and simulation are shown in Fig. 4(a), where the seed pulse duration changed from 2.7 ps (GDD, +20000 $fs_2$) to 3.0 ps (GDD, +22000 $fs_2$) and 3.4 ps (GDD, +24000 $fs_2$). The output spectrum narrowed when the seed pulse duration increased to more than 2.7 ps. This indicates that the time overlap between the pump and seed pulses improved by increasing the seed pulse duration from 1.36 to 2.7 ps; therefore, the output DC-OPA energy increased until reaching the maximum. However, when the seed pulse duration increased to more than 2.7 ps, the chirp combination exceeded the optimal matching point, causing both the output energy and the corresponding spectral bandwidth to decrease. Owing to the simulation guidance and experimental evidence, the down-chirp signal with a pulse duration of 2.7 ps was determined as the optimum match for the 3.1-ps up-chirp pump in the two-stage DC-OPA. In addition, we attempted to broaden the longer wavelength of the amplified seed pulse to 2.1 μm as calculated PM (Fig. 2b) by adjusting the time delay between the pump and seed, but both the output energy and bandwidth decreased since the intensity of the injected seed was mainly concentrated around 1.7 μm (Fig. 1a).

Next, we present the influence of the TOD of the seed pulses in the DC-OPA process. As shown in Fig. 1, a fused silica prism pair compressor was used to compress the seed pulse duration after the second stage of the DC-OPA. Here, the incident angle (56°) was determined by the Brewster's angle of the center wavelength (1.7 μm) to maintain the high throughput efficiency of the prism pair compressor. The total throughput efficiency was measured to be over 70% in this experiment. To compensate for the GDD of the seed pulses (+20,000 $fs_2$), the ideal distance separating the two prisms was determined as 531 mm (Fig. 1). This configuration of the prism pair compressor produced a TOD value of $10_5$ $fs_3$, which needed to be pre-compensated by the AOPDF to ensure the proper compression of the signal pulses. However, the injected signal pulses with such a high-order TOD made it impossible to achieve the ideal chirp matching between the pump and seed pulses in the DC-OPA process. As shown in Fig. 2(b) in the red profile, when the signal pulses contained a $10_5$ $fs_3$ TOD value, the chirp matching between the signal and pump pulses deviated from the optimal position. This caused the bandwidth of the amplified signal pulses to decrease, which was demonstrated both in the simulation and experimentally (Fig. 4b). Moreover, based on the compression results, the compressed-pulse-duration broadening caused by the narrowing of the spectrum (from 1.5–1.9 to 1.52–1.87 μm) was much smaller than that caused by the non-compensated TOD. The fused-silica prism bulk compressor was more suitable for broadband DC-OPA than the fused-silica prism pair compressor due to the low TOD value.

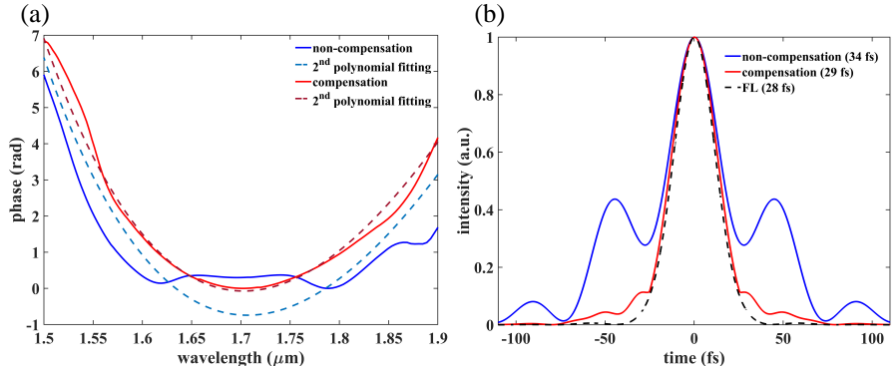

Fig. 5. (a) The simulation results of the output spectral phase caused by the DC-OPA process. Blue solid line, spectral phase corresponding to the simulated spectrum (blue solid line in the upper right corner of Fig. 4b); red solid line, spectral phase corresponding to the simulated spectrum (red solid line in the inset of Fig. 4b); blue dashed line and red dashed line are the 2nd polynomial fitting of the blue and red solid lines. (b) Simulation results of output pulse duration. Blue solid line, pulse duration corresponding to the combination of spectrum intensity (blue solid line in the inset of Fig. 4b) and spectral phase (blue solid line in Fig. 5a) with the GDD (blue dashed line in Fig. 5a) compensated; red solid line, pulse duration corresponding to the combination of spectrum intensity (red solid line in the inset of Fig. 4b) and spectral phase (red solid line in Fig. 5a) with the GDD (red dashed line in Fig. 5a) compensated; black dashed line, the transform-limited pulse duration of the spectrum (red solid line in the inset of Fig. 4b).

Finally, to achieve a shorter pulse duration of the signal pulses, we discuss the spectral phase in detail, which determines the pulse duration, and demonstrate the spectral phase distortion caused by the saturation effect in the DC-OPA process. Ross *et al.* [24] reported that the shape of the optical parametric phase (OPP) imposed by the OPA process is proportional to the wavelength-dependent phase mismatch $\Delta k$, and the OPP amplitude increases with increasing fractional pump depletion (details in Eq. (10) in Ref. [24]). As opposed to traditional stimulated-emission-based amplification, the energy flow is reversed after saturation in the OPA process. That is, the amplitude of the OPP determined by the pump depletion is reversed at the same time. Since the intensity of each wavelength injected into the DC-OPA stage is not the same (Fig. 2a), the saturation of each wavelength occurs at different times. As can be seen in the experimental and simulation results in Fig. 4(b), there was a hole in the output spectrum around the center wavelength (1.7 µm) owing to the reversed energy flow. There was a corresponding distortion in the spectral phase around the same wavelength (Fig. 5(a), blue profile). Thus, the seed pulses amplified by the DC-OPA process with an obvious saturation effect could not be compressed well by only controlling the GDD value in AOPDF (in other words, the OPP in this situation could not be fitted by a 2nd polynomial). In addition, the simulated compressed pulse shape is presented in Fig. 5(b), where there were serious pre- and post-pulses close to the main pulse, and the duration of the main pulse was much longer than the transform-limited pulse duration. To solve this problem, we added a hole (Fig. 2a) in the injected seed spectrum to void the saturation around the center wavelength. According to the simulation results presented in Figs. 4(b) and 5(a), there was no energy reversion in the amplified seed pulses, and the spectral phase distortion around the center wavelength simultaneously disappeared. The corresponding compressed pulse shape in the simulation became both much cleaner and shorter than that of the non-compensation condition, as shown in Fig. 5(b).

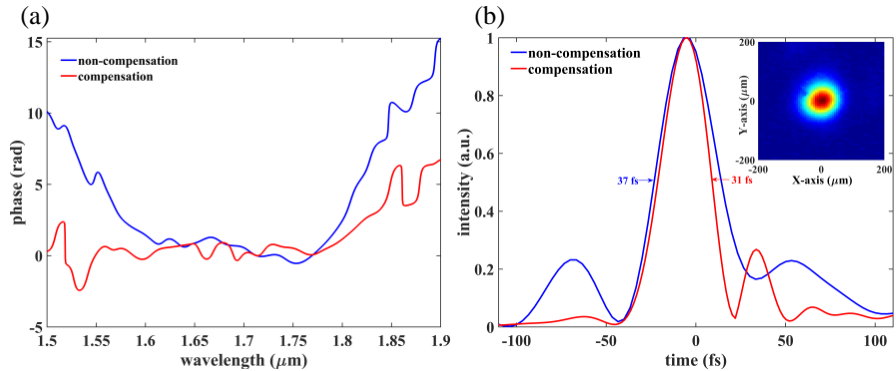

Fig. 6. (a) Experimental results of the output spectral phase after compression. Blue solid line, spectral phase corresponding to the experimental spectrum (blue solid line in Fig. 4(b)); red solid line, spectral phase corresponding to the experimental spectrum (red solid line in Fig. 4(b)). (b) Experimental results of output pulse duration. Blue solid line, pulse duration corresponding to the combination of spectral intensity (blue solid line in Fig. 4(b)) and spectral phase (blue solid line in Fig. 6(a)); red solid line, pulse duration corresponding to the combination of spectrum intensity (red solid line in Fig. 4(b)) and spectral phase (red solid line in Fig. 6(a)).

Based on the above simulation results, we performed an experiment in which a hole was added to the injected seed spectrum by the AOPDF. From the experimental spectral phase presented in Fig. 6, the final compressed pulse duration shape with a hole in the injected signal pulses was much cleaner and shorter than when saturation occurred in the DC-OPA process. The corresponding spectral phase also demonstrated this trend, where the spectral phase with compensation was much flatter than it was without compensation. Benefiting from detailed simulation guidance, we succeeded in compressing the amplified IR laser pulses to 31 fs (5.5 cycles) with a fused silica prism pair compressor. The focus spot profile of the compressed signal pulse was measured under 1.5-m focusing geometry (Fig. 6(b)), where the focal spot showed a nice intensity concentration that benefited from the low phase distortion and intensity modulation in the whole multi-TW IR DC-OPA. On the other hand, under the influence of the TOD and compensation for the saturation effect, the output pulse energy after DC-OPA decreased slightly to 98.3 mJ, which was also reflected in the difference between the spectral bandwidth (Fig. 4(b)). However, in the experiment, we were still able to obtain a signal pulse energy exceeding 100 mJ by increasing the pump intensity slightly to 82 GW/cm$_2$.

## 4. Discussion and conclusion

*4.1 Dual pump DC-OPA for generating two-cycle IR pulse*

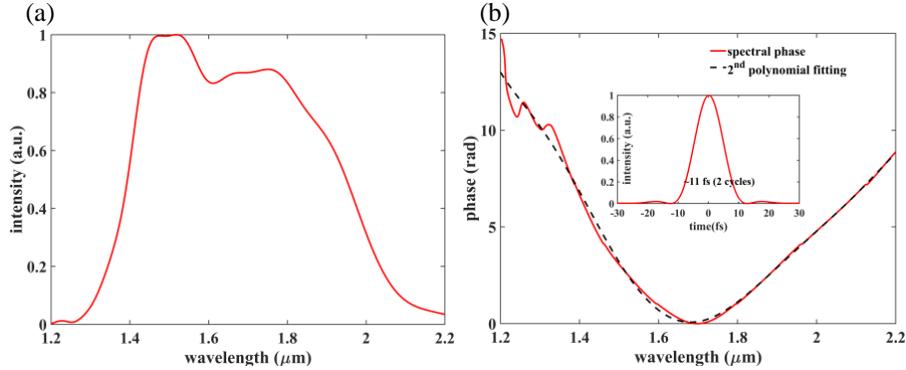

Fig. 7. (a) Simulation results of output spectrum after DC-OPA based on a dual pump DC-OPA scheme. (b) Simulation results of output spectral phase corresponding to the spectrum in Fig. 7(a) and compressed pulse duration with dispersion compensated.

To further broaden the bandwidth of the Type-I BBO DC-OPA at the center wavelength of 1.7 µm, we present a new scheme called a *dual pump* DC-OPA. As the PM efficiency in Fig. 2(b), which shows the opposite chirp matching between pump and signal used in the experiment, the seed wavelength from 1.7 to 2.2 µm was successfully amplified thanks to the optimized PM in the DC-OPA. On the other hand, the seed with a wavelength of 1.2 to 1.7 µm could be amplified using another pump pulse with the same chirp sign as the seed. This scheme was implemented by splitting the pump laser into two parts (Fig. 1). One underwent a pre-compressor and the other passed through an optical path delay to maintain the precise temporal overlap of the two pump pulses and one seed pulse in the BBO crystal. With the addition of a pre-compressor, the two pump pulses from the final grating compressor had opposite chirp signs, which is beneficial for amplifying the seed pulse in the entire spectral range. The parameters shown in Fig.1 were injected into our simulation, and the simulation results are displayed in Fig. 7, where the wavelength of 1.2 to 2.2 µm could be amplified with a well $2_{nd}$ polynomial fitting spectral phase output. In addition, we also performed a corresponding test experiment, where we changed the pump pulses to down-chirp same as the injected seed. By optimizing the time delay between the pump and seed, the injected spectrum from 1.3 to 1.7 µm was amplified (the green profile in Fig. 4(b)), which demonstrated our simulated prediction. Thus, by using a dual pump DC-OPA scheme and a bulk compressor (owing to its low TOD value) to compensate for the dispersion, we can obtain a 0.1-J IR pulse energy with a pulse duration of two optical cycles.

The present study presents the optimization of a high-energy few-cycle IR DC-OPA source employing Type-I BBO. Based on a detailed simulation analysis and experimental demonstration, a 10-Hz, 5.5-cycles, 2.3-TW IR laser pulse at a wavelength of 1.7 µm was generated. From the discussion of BBO type-I DC-OPA, a dual pump DC-OPA scheme was proposed and shown to be capable of generating a highly intense two-cycle IR pulse.

**Acknowledgment**

We thankfully acknowledge valuable discussions by Dr. O. D. Mücke. (DESY, CFEL). This work was supported in part by the Ministry of Education, Culture, Sports, Science, and Technology of Japan (MEXT) through Grant-in-Aid under Grant (17H01067, 19H05628), in part by the MEXT Quantum Leap Flagship Program (Q-LEAP) Grant Number JP-MXS0118068681, in part by the FY 2019 President discretionary funds of RIKEN, and in part by the Matsuo Foundation 2018.